\pdfoutput=1
\documentclass[a4paper,referee]{raa}            % referee version: for submission
\usepackage{edtable}

\let\oldequation\equation
\let\oldendequation\endequation

\renewenvironment{equation}
{\linenomathNonumbers\oldequation}
{\oldendequation\endlinenomath}

\usepackage{orcidlink}
\usepackage{graphicx,times}             %for PS/EPS graphics inclusion, new
\usepackage{natbib}
\usepackage{amssymb,amsmath}
\usepackage{subcaption}
\usepackage{bm}
\bibpunct{(}{)}{;}{a}{}{,}
\usepackage{lineno}
\let\oldflalign\flalign \let\oldendflalign\endflalign  \renewenvironment{flalign}    {\linenomathNonumbers\oldflalign}    {\oldendflalign\endlinenomath}
\begin{document}
  \title{The {Mini-SiTian} Array: Prospects of Searching for Tidal Disruption Events 
}
%   \subtitle{I. Place Your Subtitle Here}

   \volnopage{Vol.0 (20xx) No.0, 000--000}      %%preserved for Editor. DOn't remove!
   \setcounter{page}{1}          %%starting page, preserved for Editor. DOn't remove!

   \author{Bei-chuan Wang~\orcidlink{0009-0001-5324-2631}   %% Put your Chinese name in "( )" if you like. Note to open line 11 "\usepackage[UTF8]{ctex}"
      \inst{1,2},
    Jun-jie Jin~\orcidlink{0000-0002-8402-3722}
      \inst{1},
      Yu Zhang
      \inst{1},
      Yanan Wang
      \inst{1},
      Song Wang
      \inst{1,3},
      Hong-rui Gu~\orcidlink{0009-0007-5610-6495}
      \inst{1,2},
      Min He
      \inst{1},
      Hai-yang Mu
      \inst{1},
      Kai Xiao~\orcidlink{0000-0001-8424-1079}
      \inst{1,3},
      Zhi-rui Li
      \inst{1,2},
      Zhou Fan
      \inst{1,2},
      Liang Ge
      \inst{1,2},
      Jian-feng Tian
      \inst{1,2},
      Yang Huang
      \inst{1,2},
      Jie Zheng~\orcidlink{0000-0001-6637-6973}
      \inst{1},
      Hong Wu
      \inst{1,2} and
      Jifeng Liu
      \inst{1,2}}
%% Here is an example of three authors come from different institutes.
%% For single author or all the authors from an institute, use "\inst{}" only

   \institute{Key Laboratory of Optical Astronomy, National Astronomical Observatories, Chinese Academy of Sciences, Beijing 100101, People's Republic of China; 
   {\it wangbc@bao.ac.cn, jjjin@bao.ac.cn}\\
%% Please give the E-mail address of the author, to whom future correspondence and
%% offprint requests will be sent.
        \and
        School of Astronomy and Space Science, University of Chinese Academy of Sciences, Beijing 100049, People's Republic of China
        \and 
        Institute for Frontiers in Astronomy and Astrophysics, Beijing Normal University, Beijing, 102206, China\\
\vs\no
   {\small Received 20xx month day; accepted 20xx month day}}

\abstract{We assess the detectability of tidal disruption events (TDEs) using mock observations from the {Mini-SiTian} array. We select 100 host galaxy samples from a simulated galaxy catalog based on specific criteria such as redshift, BH mass, and event rate. Taking into account the site conditions and survey strategy, we simulate observations over a 440 deg$^2$ field. The results indicate that $0.53\pm 0.73$ TDEs can be detected per year when observing in both $g$ and $r$ bands with 300-second exposures every 3 days. Applying this method to the SiTian project, we expect to discover approximately 204 TDEs annually, heralding a new era in TDE science.
}

   \authorrunning{Wang et al.}            %author_head in even pages
   \titlerunning{The {Mini-SiTian} Array: Prospects of Searching for Tidal Disruption Events}  % title_head in odd pages

   \maketitle
%% The author head (on even pages) and the title head (on odd pages) will be
%% automatically extracted from \author{} and \title{}. Whenever the title is too long,
%% you will be asked to supply a shorter one by inserting either \authorrunning{} or
%% \titlerunning{} before \maketitle. Anyway, you can specify your own heads.
%%
%%
%% Note: In the following text body of your manuscript, please note several differences from
%%       other major journals:
%% (1) \subsection{Please Capitalize the First Letter of Each Notional Word in Subsection Title}
%% (2) Please Capitalize the First Letter of Each Notional Word in all s' captions
%
%________________________________________________ sections below
\section{Introduction}           
%% first-level sections will be auto-capitalized
\label{intro}
\subsection{Tidal disruption event}
\label{TDEs}
% Introduce the Basic Concept of TDEs

When a star gets close to the massive black hole (BH) at the center of a galaxy, it can be torn apart if the tidal force by the BH overpowers the self-gravity of the star (\citealt{Rees1988,Evans1989}). After the disruption, approximately half of the star's mass escapes from the BH, while the remaining half of stellar debris moves in a highly eccentric orbit around the BH and then can be accreted by the BH to form an accretion disk. 
%Under certain mechanisms, a 
A luminous flare of electromagnetic radiation is produced in this process, lasting for months to years.
Such a phenomenon, named as tidal disruption event (TDE), can be visible from radio to X-ray wavelengths.

TDEs have aroused extensive interests because of their distinctive scientific merits. Primarily, TDEs provide compelling evidence for the presence of supermassive black holes (SMBHs) in normal galaxies, especially in dwarf galaxies and distant quiescent galaxies (\citealt{Maguire2020}). TDEs are also useful in searching for specific types of BHs, such as intermediate-mass black holes (IMBHs) and binary black holes (BBHs), which are generally quite difficult to detect (\citealt{French+etal+2020,Greene2020,Shu+etal+2020,Huang2021,Huang+etal+2024}). Moreover, TDEs serve as an ideal laboratory for understanding the accretion physics of SMBHs by monitoring the entire process of BH activity and even by witnessing the formation and evolution of jets. TDEs also provide a new method to probe the sub-parsec environment of the quiescent SMBHs through observations of the infrared and radio echoes (\citealt{Gezari2021,Jiang2021,VanVelzen2021}). Finally, in the era of multi-messenger astrophysics, TDE is thought as an important astrophysical process for producing high-energy neutrinos (\citealt{Stein+2021}).

TDEs were first identified as soft X-ray transients in galactic nuclei based on ROSAT archive data in the late 1990s (\citealt{Bade1996,Komossa1999}). Subsequent X-ray instruments, such as $XMM$–$Newton$ , $Chandra$ , and $Swift$ , have also discovered a few TDE candidates (\citealt{Weisskopf+etal+2000,Gehrels+etal+2004,Saxton+etal+2008}). 
Recently, optical photometric surveys open up new avenues for the detection of TDEs (\citealt{vanVelzen2020,Gezari2021}). 
With the operation of surveys such as the Panoramic Survey Telescope And Rapid Response System (Pan-STARRS), the Palomar Transient Factory (PTF), the All-Sky Automated Survey for Supernovae (ASAS-SN), and most notably, the Zwicky Transient Facility (ZTF), the discovery number of TDEs has significantly increased to $>10$ events per year (\citealt{Gezari2012,Arcavi2014,Holoien2016,Gezari2021,Hammerstein2023,Yao+etal+2023}). 
Despite these advancements, the TDE rate remains exceedingly low, typically on the order of $10^{-4}-10^{-5}$ per galaxy per year for most galaxies (\citealt{Gezari2008,VanVelzen2014,Holoien2016,Velzen2018,Yao+etal+2023}). 
This rarity emphasizes the importance of large-scale surveys, such as the 
upcoming Vera C. Rubin Observatory (VRO) in the southern hemisphere, which will be crucial for expanding the TDE sample size. Different researchers predicted that at least thousands of TDEs ($\sim$ $1600-8000$) per year can be discovered by VRO (\citealt{Ivezic+etal+2019,Bricman2020,Roth2021}). 
However, all these surveys (e.g., ZTF, VRO) are designed with an observational cadence of days, which means much detailed information during the TDE process would be missed.
Especially, the timescale of the rising phase of TDEs is particularly short and may include essential information about disruption processes. Therefore, a time-domain survey with large field of view, deep limiting magnitude and high cadence is urgent for TDE studies, especially scans in the northern hemisphere to complement VRO in sky coverage.
\subsection{The SiTian project and {Mini-SiTian}}
\label{SiTian}
\begin{figure}[htbp]
   \centering
  \includegraphics[width=\textwidth, angle=0]{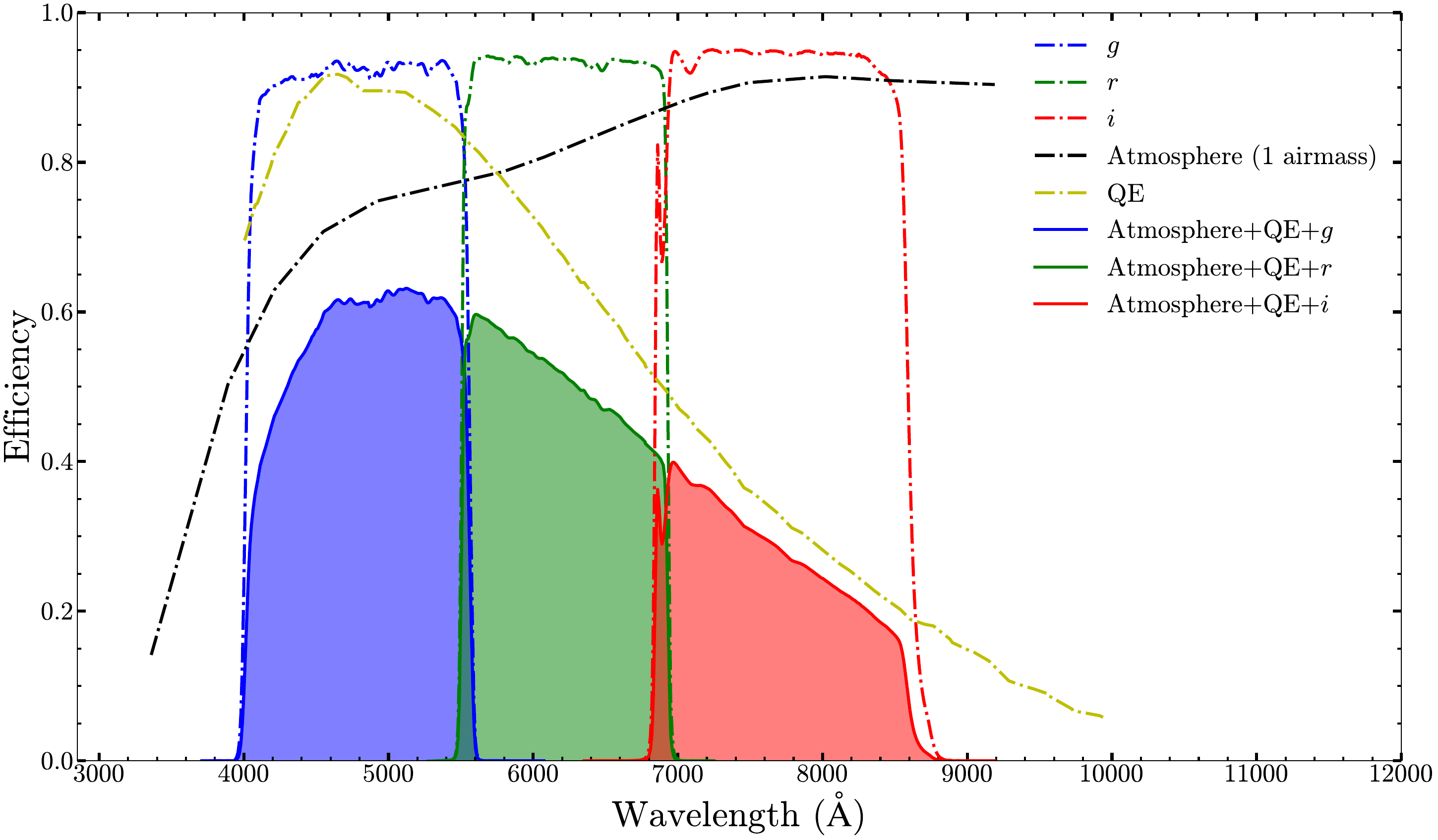}
   \caption{The throughput curves for the $g$ filter (blue dotted-dashed line), $r$ filter (green dotted-dashed line), $i$ filter (red dotted-dashed line), along with the atmosphere (black dotted-dashed line), and the quantum efficiency (QE) of the CMOSs (yellow dotted-dashed line), are shown. The total throughput curve for each {Mini-SiTian} filter is represented by a solid line in the corresponding color.} 
   \label{fig1}
   \end{figure}
The SiTian project offers an opportunity to address these challenges.
SiTian is a ground-based optical monitoring project, proposed by \citet{Liu2021}. It has 20 groups of 1-m-class telescopes and every group includes 3 telescopes equipped with $g,r,i$ filters. The field of view (FOV) of each telescope is $5^{\circ}\times 5^{\circ}$. The optical design of each telescope has been designed utilizing a catadioptric system, an evolution of the modified Schmidt telescope, designed to optimize the performance of the telescope array for its science goals (\citealt{Chen2022}). SiTian will scan at least 10,000 deg$^2$ of the sky every half hour, down to a detection limit of $V \approx 21$ mag with three filters operating simultaneously. 
{It will discover numbers of new celestial objects and phenomena, and is expected to achieve breakthroughs in the study of extreme high-energy explosive sources, electromagnetic counterparts of gravitational waves, exoplanets and celestial bodies in the solar system.} The SiTian project will play an important role in the research of major scientific issues, such as dark matter, dark energy, and black holes.

% compare ZTF with 1-m SiTian
The SiTian project has its own unique advantages, especially in the detection of optical TDEs, compared to other transient surveys. Unlike ZTF, SiTian's key strength in studying TDEs lies in its ability to observe three bands simultaneously, allowing the use of multi-band color evolution to identify TDE candidates. Moreover, the cadence of SiTian is 80 times higher than that of ZTF, allowing for a stacked image depth down to 22.35 magnitude within a single night (\citealt{Liu2021}). The combination of higher cadence and deeper photometry from stacked images is crucial for detecting the rising phase of TDEs and determining their position within galaxies, which are critical for the early identification of TDE candidates and reducing the pressure on subsequent spectroscopic observations. 
Compared to VRO, SiTian has the advantage of obtaining better-sampled light curves for TDEs. While VRO is capable of capturing transients with a magnitude of $\sim 24.5$ in the $r$ band within 30 seconds, it covers the sky with a cadence of \textit{three days}, resulting in approximately 800 observations over the main survey areas of 10 years (\citealt{Ivezic+etal+2019}). This relatively low cadence makes it challenging to obtain a complete light curve for transients like TDEs, particularly in capturing the rising phase. Additionally, VRO will mainly scan the southern celestial sphere. While SiTian covers the northern celestial sphere, complementing each other to achieve all-sky monitoring.

\begin{figure}[htbp]
   \centering
  \includegraphics[width=\textwidth, angle=0]{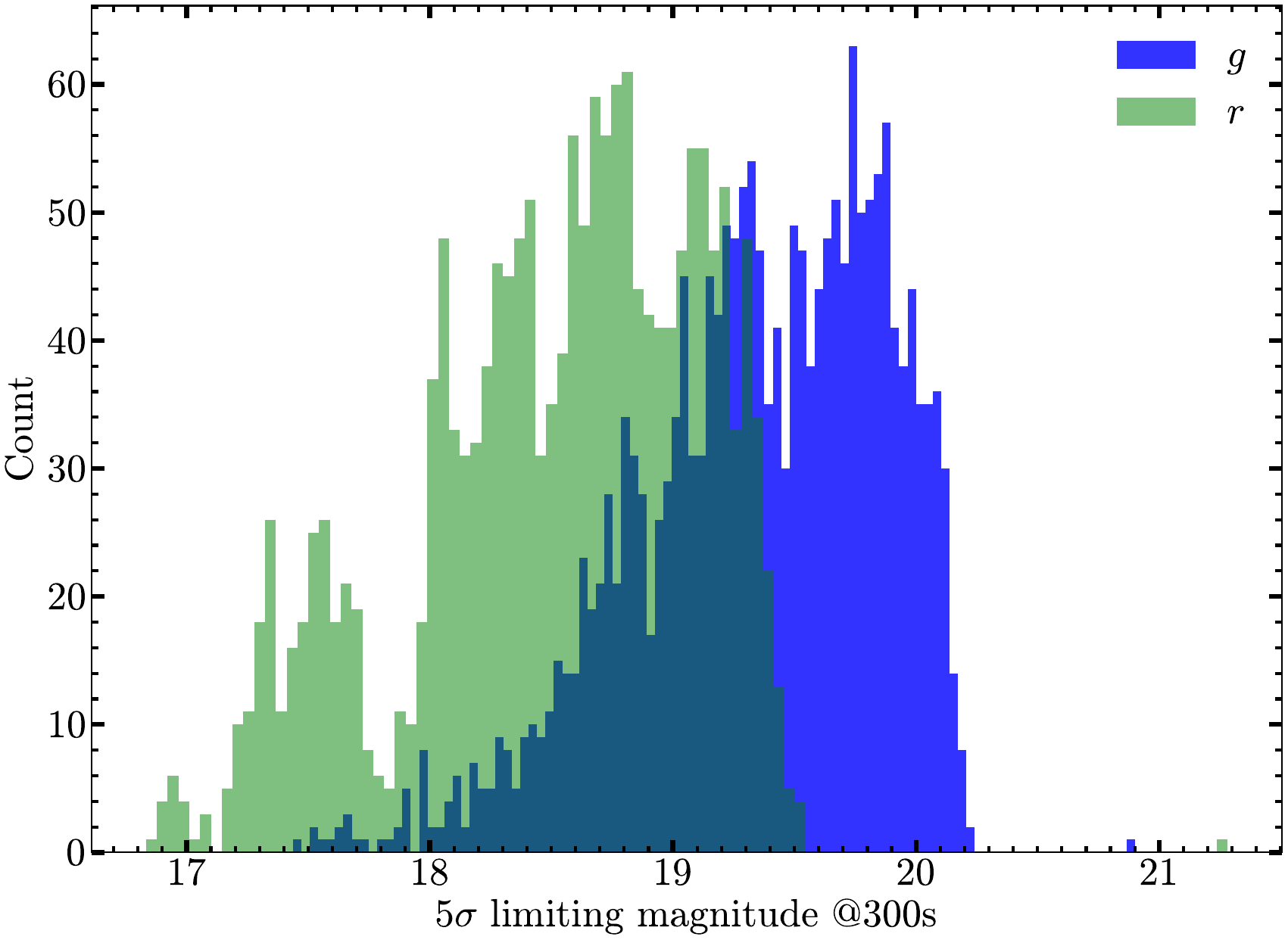}
   \caption{The histogram of the 5$\sigma$ limiting magnitudes of $g$ band (filled with blue) and $r$ band (filled with green) of {Mini-SiTian} when the exposure time is 300 s.} 
   \label{fig2}
   \end{figure}
The {Mini-SiTian} array, a compact and smaller version of the SiTian array, acts as a pathfinder to test the control systems, data processing pipelines, and the feasibility of the technology and observing strategies of SiTian (See M. He et al. 2025). 
It is located at the Xinglong Observatory and has started operations in 2022. The array consists of three 30-cm in diameter telescopes equipped with $g$, $r$, $i$ filters, respectively.
Each telescope has an effective field of view of $2.29^{\circ}\times 1.53^{\circ}$, by mounting a 9K $\times$ 6K scientific Complementary Metal-Oxide-Semiconductor (CMOS) detector on the back of the telescope.
Figure \ref{fig1} shows the transmission curves of the filters and the total throughput of {Mini-SiTian} in the three bands (He et al. 2025).
The top-level telescope specifications are listed in Table \ref{tab1}. Figure \ref{fig2} shows limiting magnitudes obtained from the {Mini-SiTian} historical data, illustrating the statistical distributions of the 5-sigma limiting magnitudes in the $g$ and $r$ bands with a 300-second exposure. These distributions will be utilized in Section \ref{Mockobs} to derive the varied daily limiting magnitudes. The current scientific goals of {Mini-SiTian} include a northern survey for the variable objects and follow-up observations of new transients like Gravitational Wave (GW), comets, supernovae, and TDEs (See H.G. Han et al. 2025, K. Xiao et al. 2025).
\begin{table}
\bc
\centering
%\begin{minipage}[]{150mm}
\caption[]{Top-level specifications for the {Mini-SiTian} array. (M. He et al. 2025).
\label{tab1}}%\end{minipage}
\setlength{\tabcolsep}{1pt}
\small
 \begin{tabular}{ll}
  \hline\hline\noalign{\smallskip}
\multicolumn{1}{c}{Item} &  \multicolumn{1}{c}{Specifications}\\
  \hline\noalign{\smallskip}
% new variable
Aperture              & 0.3 m   \\
Focal ratio          & f/3                 \\
Camera FOV         & $2.29^{\circ}\times 1.53^{\circ}$                       \\
Filters            & $g$, $r$, $i$\\
Pixel scale           & 0.862 arcsec per pixel \\
Pixel size            &3.76 $\mu$m \\
Limiting magnitude      &$m_{g,\rm lim} = 19.5, m_{r,\rm lim} = 19.0$ for 300-s exposures\\
  \noalign{\smallskip}\hline
\end{tabular}
\ec
%% place \tablecomments and \tablerefs below \end{center| and \end{center}:
%% you may leave the table-width parameter to editors or set to your actual size
\end{table}

In this work, we focus on evaluating {Mini-SiTian}'s capability to detect TDEs using mock observations.
The spectra of the optical-ultraviolet TDEs are dominated by a strong and consistent blue continuum (\citealt{vanVelzen2020}). Given the importance of color information for identifying TDEs and the total throughput of the filters, the $g$ and $r$ bands are used in our simulations.

This paper is organized as follows. 
Section \ref{sample} describes the galaxy samples used for modeling TDEs, followed by the modeling methods in Section \ref{method}. Section \ref{result} presents the main results from the mock observations, with a comparison to the observation of optical TDEs discussed in Section \ref{discussion}.
Finally, Section \ref{summary} provides a brief summary.
%% Authors can give a citation as 'Michel et al. 1992'.
%% You may also use \cite, \citep and \citet for citation, and use ~1 or Figure~1
%% and so forth. Using \ref and \label for cross-references of s/Figures
%% is a good way in adjusting/adding/removing text, s or figures.

\section{Galaxy Sample for TDE modeling}
\label{sample}
In the standard theory of TDE (\citealt{Rees1988}), the disrupted stellar materials quickly circularize and form an accretion disk. Following the disruption, the light curve is predicted to follow a $t^{-5/3}$ decay, driven by the mass fallback rate. However, observations of optical TDEs frequently deviate from the $t^{-5/3}$ decay pattern \citep{Gezari2021}.
Several empirical tools have been developed to produce TDEs' light curves by simulating their physical processes.

Among these tools, {\tt MOSFiT} (\citealt{Guillochon2018}) is capable of not only fitting observational data but also producing light curves based on model predictions. By using FLASH simulations of TDEs, {\tt MOSFiT} integrates components such as the {\it Fallback Engine}, {\it Viscous Delay}, and {\it Photosphere} modules (\citealt{Mockler2019}). The comprehensive approach allows {\tt MOSFiT} to provide an accurate and statistically robust representation of TDE light curves, thereby bridging the gap between theoretical expectations and empirical observations.

Because observational galaxy catalogs lack the parameters required by the {\tt MOSFiT} model and are incomplete, we choose a large synthetic galaxy catalog, CosmoDC2 (\citealt{Korytov2019}), to model TDEs. 
This catalog includes $\sim$ 2.26 billion galaxies covering a sky area of 440 deg$^2$, extending to a redshift of $z=3$.
It matches the expected number densities from current surveys down to a magnitude depth of $m_r=28$. 
The redshift $z$, the total stellar mass $M_{\rm TSM}$, and the central BH mass $M_{\rm BH}$ for each galaxy are extracted for modeling TDE light curves with {\tt MOSFiT}.

Given the large size of the catalog and flux limit of the photometric survey, we first excluded galaxies outside the parameter spaces where TDEs can occur. The following three criteria were adopted:

(i) Excluding galaxies with $z>1$, as the luminosity distribution of detected optical TDEs indicates that both {Mini-SiTian} and SiTian cannot reach such distance. TDEs at such distance are too faint to be observed.

(ii) Excluding galaxies with $M_{\rm BH}<10^5 M_{\odot}$. Theoretically, TDEs can occur in such galaxies, e.g., white dwarf disruptions by IMBHs (see review by \citealt{Maguire2020}). However, their physical details and corresponding optical light curves remain unclear. Additionally, such TDEs may be difficult to detect with {Mini-SiTian} due to their dim luminosities.

(iii) Excluding galaxies with $M_{\rm BH}>M_{\rm Hills}$. The Hills mass, denoted as  $M_{\text{Hills}}$, represents the upper limit on the mass of a Schwarzschild black hole to ensure that its event horizon is within the tidal radius ($R_t\equiv(M_{\rm BH}/M_{\ast})^{1/3}R_\ast$, where $M_\ast$ and $R_\ast$ are the mass and radius of a star, respectively).
It is defined as (\citealt{Hills1975,Leloudas2016})
\begin{flalign}
\label{eq1}
&\ M_{\rm Hills}=9\times10^7M_{\odot}(\frac{M_{\ast}}{M_{\odot}})^{-1/2}(\frac{R_\ast}{R_{\odot}})^{3/2}.&
\end{flalign}
We use a Kroupa initial mass function (IMF) for stellar populations \citep{Kroupa+etal+1993}, with minimum and maximum stellar masses of 0.08 and 1 $M_{\odot}$, respectively. The relation $R_{\ast} \propto M^{0.8}_{\ast}$ on the lower main sequence is applied to obtain $M_{\rm Hills} \propto M^{0.7}_{\ast}$ \citep{Stone2016}. For stars with a mass comparable to the solar mass, we determine the Hills mass to be $M_{\rm Hills} = 9.0 \times 10^7 M_{\odot}$. Thus, galaxies with black holes of $M_{\rm BH} > 9.0 \times 10^7 M_{\odot}$ are excluded.

Applying the above criteria, we derive a parent sample consisting of $\sim$ 22,600,000 galaxies.
However, given the low TDE rate, it is unlikely that TDEs will occur in most of these galaxies over a year.
Therefore, we focus only on the galaxies where TDEs are likely to happen.
Here we adopt the TDE rate provided by  \citet{Stone2016} with the modification of cosmological effect, that is
\begin{flalign}
&\ \dot{N}_{\rm TDE}(z)=\dot{N}_{\rm TDE,0}\frac{\mathrm{d} t_0}{\mathrm{d} t}=10^{-4.19}\frac{1}{1+z}(\frac{M_{\rm BH}}{10^8M_{\odot}})^{-0.223} {\rm yr}^{-1},&   
\end{flalign}
where $t_0$ and $t$ are the time in the rest-frame and observer-frame, respectively. 
For each of these $\sim$ 22,600,000 galaxies, we calculate $\dot{N}_{\rm TDE}(z)$ and generate a random number between 0 and 1.
If $\dot{N}_{\rm TDE}(z)$ is larger than the random number, a TDE is assumed to occur in that galaxy.
We repeat this process 100 times to reduce the influence of randomness. This yields an average number of host galaxies with TDEs $N_{\rm host} = 2478 \pm 48$, corresponding to an event rate of approximately $\sim 0.011 \%$. These selected galaxies will be used in the following analysis.
\section{Method}
\label{method}
To estimate the detection rate of TDEs by {Mini-SiTian}, we simulate the light curves of TDEs based on their physical properties. Considering the observation conditions at Xinglong Observatory, we can measure the number of TDEs detected per year.
This section outlines the steps of our method.

\subsection{TDE light curves generated by {\tt MOSFiT}}
\label{MOSFiT}

\begin{table}
\bc
\centering
\caption[]{Summary of parameters used in {\tt MOSFiT} for modeling TDEs.
\label{tab2}}
\setlength{\tabcolsep}{1pt}
\small
 \begin{tabular}{lllll}
  \hline\hline\noalign{\smallskip}
\multicolumn{1}{c}{Parameter} & \multicolumn{1}{c}{Description} &\multicolumn{1}{c}{Value} & \multicolumn{1}{c}{Units} & \multicolumn{1}{c}{Reference}\\
  \hline\noalign{\smallskip}
$z$&redshift& {0-1} & &  --               \\ 
$M_{\rm BH}$ &black hole mass        & $1\times 10^5-9\times 10^7$ & $M_{\odot}$&--     \\
$M_{\ast}$ &stellar mass          & 0.08-1                         &  $M_{\odot}$ &  \citet{Kroupa+etal+1993,Lin2022}  \\
$n_{\rm H,host}$&hydrogen column density     & $0-4.3\times 10^{21}$       &  ${\rm cm}^{-2}$   & \citet{Garn2010,GonzalezDelgado2015}             \\
$b$& scaled impact parameter                 & 0-2                       &        & \citet{Bricman2020,Lin2022}         \\
$\epsilon$&radiative efficiency          & 0.1                       &      &\citet{Mockler2019}   \\
$l_{\rm ph}$&photosphere power-law exponent         & 1.5                       &       &\citet{Mockler2019}           \\
$R_{\rm ph,0}$&photosphere power-law constant       &6.3                   &       &\citet{Mockler2019}           \\
$\sigma$&model variance             &0.1                        &     &\citet{Mockler2019}             \\
$T_{\rm viscous}$&viscous delay time scale    &0.001–0.5                  &      days   &\citet{Mockler2019}     \\
$t_{\rm 0}$&days since first detection          &0                          &      days  & --       \\
  \noalign{\smallskip}\hline
\end{tabular}
\ec
%% place \tablecomments and \tablerefs below \end{center| and \end{center}:
%% you may leave the table-width parameter to editors or set to your actual size
\end{table}

To model the TDE light curves with {\tt MOSFiT}, 11 prior parameters are used in the configuration files. 
Below we provide a detailed discussion on the treatment of some key parameters. In {\tt MOSFiT}, $n_{\rm H,host}$ is used to calculate extinction following the relation $n_{\rm H,host}=1.8\times10^{21}A_V$. Although CosmoDC2 has provided dust extinction parameters $A_V$ and $R_V$, both parameters are too low to describe the extinction at the centers of host galaxies accurately. Instead, we calculate $A_{\rm H\alpha}$ from $M_{\rm TSM}$ of the host galaxy (\citealt{Roth2021}) and convert it to $A_V$ (\citealt{Calzetti2000}). Specifically, galaxies are classified into star-forming and quiescent galaxies through specific star formation rate. For star-forming (SF) galaxies which have a specific star formation rate higher than $10^{-11.3}$ ${\rm yr}^{-1}$, $A_{\rm H\alpha}$ is sampled from a Gaussian distribution with a minimum value of zero. The median is calculated by
\begin{flalign}
&\ A_{\rm H\alpha, median}=0.91+0.77x+0.11x^2-0.09x^3, &  
\end{flalign} 
where $x\equiv \log_{10}(\frac{M_{\rm TSM}}{10^{10}M_{\odot}})$, with a standard deviation of 0.28 mag. This relationship is only calibrated for SF galaxies with $M_{\rm TSM}$ ranging between $10^{8.5}$ and $10^{11.5}M_{\odot}$ (\citealt{Garn2010}). For the galaxies outside this range, the edge values are used. For quiescent galaxies which have a specific star formation rate lower than $10^{-11.3}$ ${\rm yr}^{-1}$, $A_{\rm H\alpha}$ is sampled from a Gaussian distribution with a minimum value of zero, where the median and the standard deviation are 0.2 and 0.06 mag, respectively. The choice of the median extinction at the galaxy center is based on the results of \citet{GonzalezDelgado2015}. We then adopt the law from \citet{Calzetti2000} and $R_V = 4.2$ to convert all $A_{\rm H\alpha}$ into $A_V$. Finally, the dust extinctions from the host galaxy and the Milky Way are both applied to each event according to the model in \citet{O'Donnell1994}.

The parameter $b$ is related to the impact factor $\beta$ ($\beta \equiv R_t/R_p$, where $R_p$ is the pericenter radius). Following \citet{Bricman2020}, we randomly assign a $\beta$ value to each disruption event from the distribution function of $\beta$,
\begin{flalign}
&\ P(\beta)=\frac{2}{\beta^3}(\frac{1}{\beta_{\rm min}^2}-\frac{1}{\beta_{\rm max}^2})^{-1}, &
\end{flalign}
where 
\begin{flalign}
&\ \beta_{\rm min}=\left\lbrace
        \begin{aligned}
            &0.5 ,& (0.08\leqslant M_{\ast}/M_{\odot}\leqslant 0.3)\\
            &0.5+0.1(\frac{(M_{\ast}/M_{\odot})-0.3}{1.0-0.3})  ,& (0.3<M_{\ast}/M_{\odot}\leqslant1.0)\\
        \end{aligned}\right. &
\end{flalign}
\begin{flalign}
&\ \beta_{\rm max}=11.8(\frac{M_{\rm BH}}{10^6M_{\odot}})^{-\frac{2}{3}}(\frac{M_{\ast}}{M_{\odot}})^{\frac{7}{15}}. &
\end{flalign}
The selection of $\beta$ will be explained below. $\beta$ is then converted into $b$ following the method described in \citet{Lin2022}. The {\tt MOSFIT} TDE model employs a hybrid polytropic approach that interpolates between polytropic indices $\gamma = 5/3$ when $M_{\ast}/M_{\odot} \leqslant 0.3$ and $\gamma = 4/3$ when
$M_{\ast}/M_{\odot} \geqslant 1$, where the fraction corresponding to $\gamma = 5/3$ is defined as
\begin{flalign}
&\        f=\left\lbrace
        \begin{aligned}
            &1 ,& (0.08\leqslant M_{\ast}/M_{\odot}\leqslant 0.3)\\
            &1-\frac{(M_{\ast}/M_{\odot})-0.3}{1.0-0.3}  ,& (0.3<M_{\ast}/M_{\odot}\leqslant1.0)\\
        \end{aligned}\right. &
\end{flalign}
within the mass range from which we drew \citet{Mockler2019}. Therefore, for a star with $0.08\leqslant M_{\ast}/M_{\odot} \leqslant1.0$,
\begin{flalign}
&\        \beta=\left\lbrace
        \begin{aligned}
            &(1.2-0.8f)b+0.6-0.1f ,& (0\leqslant b\leqslant 1)\\
            &(2.2-0.6f)b-0.4-0.3f  .& (1<b\leqslant2)\\
        \end{aligned}\right.&
\end{flalign}

\begin{figure}[htbp]
   \centering
  \includegraphics[width=\textwidth, angle=0]{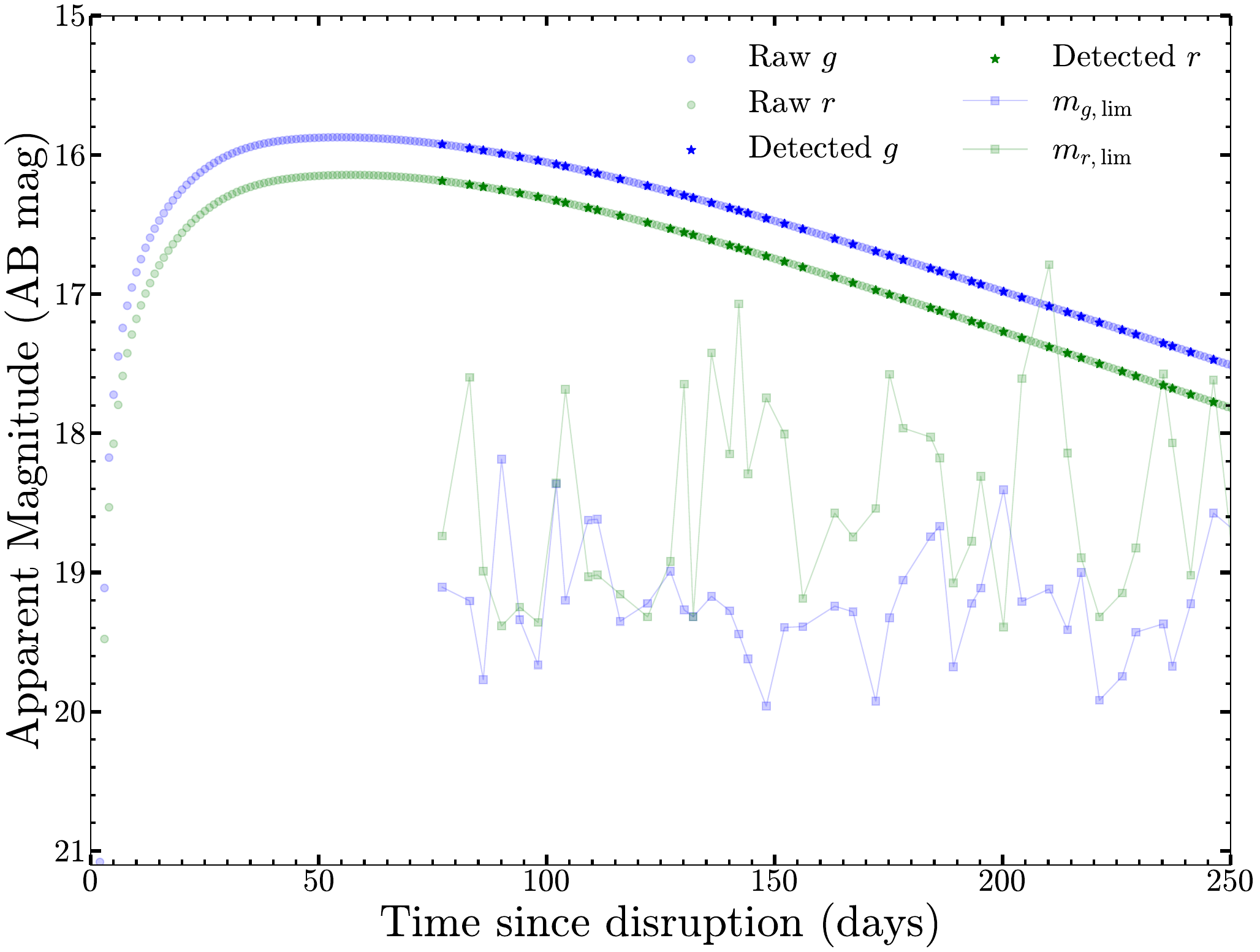}
   \caption{An example of TDE light curves (blue represents the $g$ band and green represents the $r$ band) of our mock observations. The adopted parameters are: $M_{\rm BH}=2.2\times 10^6 $ $M_{\odot}$, $M_{\ast}=0.289$  $M_{\odot}$, $z=0.032$, $b=0.706$, and $n_{\rm H,host}=5.3 \times 10^{20}$ ${\rm cm}^{-2}$. The raw light curve generated by {\tt MOSFiT} (represented by dots) has been sampled from day 1 to day 250 after the disruption, with an interval of 1 day. The data for the detected TDE (represented by stars) and limiting magnitudes at the $g$ and $r$ bands (connecting lines) have also been plotted.} 
   \label{fig3}
   \end{figure}

The $\beta$ is converted to $b$, using a piecewise linear function $b=b(\beta)$, with an upper limit of $b_{\rm max}=2$.
We obtain $z$ and $M_{\rm BH}$ values from the galaxy catalog. For $\epsilon$ (radiative efficiency), $l_{\rm ph}$ (photosphere power-law exponent), $R_{\rm ph,0}$ (photosphere power-law constant), $\sigma$ (model variance), and $T_{\rm viscous}$ (viscous delay time scale), we adopt their values from \citet{Mockler2019}. All parameters used in {\tt MOSFiT} for modeling TDEs are summarized in Table \ref{tab2}. With these parameters and the {Mini-SiTian} filter throughput curves, we derive the TDE light curves at {the $g$ and $r$ bands} (see an example in {Figure \ref{fig3}}).
\subsection{Mock Observations}
\label{Mockobs}
The steps for creating mock observations are as follows. First, a specific source can only be optimally observed for half-time in a year, so we put ``observation windows'' spanning 180 days within the timeline. Second, we consider the weather conditions at the Xinglong Observatory. We define an observable night as one with observing time exceeding 4 hours. According to past weather data, approximately $61\%$ of the nights are observable each year (\citealt{Zhang+etal+2015}; M. He et al. 2025). For simplicity, we use a parameter $P_{\rm obs}=61\%$ to describe the probability that a night is observable. Third, we assume a simple uniform survey strategy for the experimental 440-{deg$^2$} CosmoDC2 field, which will be scanned with a 300-s single exposure every $n$ days using $g$ and $r$ bands. Ideally, $n$ is set to 3, but when considering weather, $n$ is sampled from a Gaussian distribution with a minimum value of 2 and a maximum value of 6, where the median and the standard deviation are 3 and 1, respectively. Fourth, we require that the host galaxies of TDEs are distinctly detected in the $g$ or $r$ band of {Mini-SiTian}, which means the TDEs contrast sufficiently against the light from the host galaxies. Specifically, for each detected TDE, we impose the following criteria: $m_{i,\rm host}<m_{i,\rm lim}$ and $m_{i,\rm peak}<m_{i,\rm host}+\Delta m$, where $i$ represents either $g$ or $r$ band, $m_{i,\rm host}$ is magnitude of host galaxy at $i$ band, $m_{i,\rm peak}$ is the peak magnitude of a TDE at $i$ band, $\Delta m$ is the survey contrast quantity, indicates how easily a survey can detect faint flares against the host light, with larger values corresponding to better detection capabilities. We apply a more typical practical limit $\Delta m=1$ used in \citealt{Roth2021}. Lastly, to identify a TDE candidate only based on the {Mini-SiTian} data, it is crucial to consider the number of high-quality data points in the light curve, specifically, how many times a TDE is detected above the limiting magnitude in both the $g$ and $r$ bands. We randomly derive the daily limiting magnitudes based on the distribution of historical observational data, allowing our simulations close to real observational conditions. Here we define a detection as having at least ten observed data points brighter than the limiting magnitude in both the $g$ and $r$ bands, consistent with the criteria used by \citet{Lin2022} and \citet{Bricman2020}. This criterion not only reduces contamination from fake sources and moving objects in real observations, but also obtains the trend of TDE light curves, allowing for improving observational strategies. At last, we model a total of 100 groups of TDEs from selected galaxies (see Section \ref{sample}). An example of a detected TDE is shown in Figure \ref{fig3}.
 
\section{Results}
\label{result}
\begin{figure}[htbp]
   \centering
  \includegraphics[width=\textwidth, angle=0]{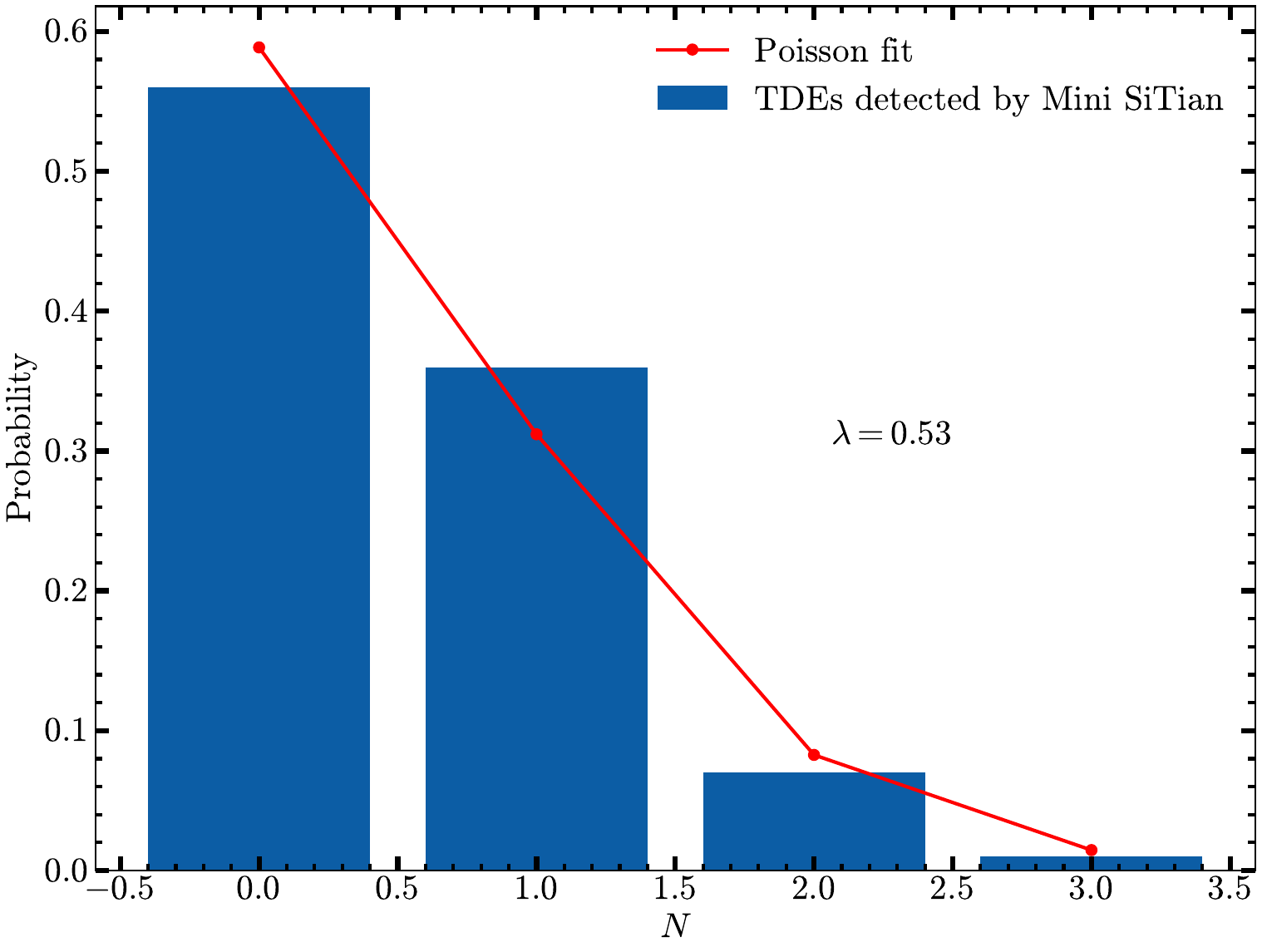}
   \caption{The number distribution of TDEs detected by {Mini-SiTian} for 100 group samples. The total number of detected TDEs for the 100 group samples is 53. $\lambda$ represents the mean of the Poisson function.}
   \label{fig4}
   \end{figure}
In this section, we will present the results of TDE detection rate from the {Mini-SiTian} mock observations and estimate the TDE detection rate for the SiTian project.

Figure \ref{fig4} illustrates the number distribution of the TDEs ($N$) detected by {Mini-SiTian} under the 100 group samples. The number of TDEs detected by {Mini-SiTian} varies from 0 to 3, depending on the sample size of selected galaxies.
We adopt a Poisson function to fit the distribution and obtain the mean of $\lambda=0.53$ and standard deviation of $\sqrt{\lambda}=0.73$. Overall, the {Mini-SiTian} array can discover approximately $0.53\pm 0.73$ TDE candidates per year, observed in the $g$ and $r$ filters within the CosmoDC2 field
at a uniform cadence of approximately 3 days. TDE candidates almost have $m_{r,\rm peak}<19$.
This would enable follow-up spectroscopic observations using telescopes with apertures of 2 meters, for instance, the 2.16-m Telescope (\citealt{Fan2016}) located at Xinglong Observatory.
\begin{figure}[htbp]
   \centering
  \includegraphics[width=\textwidth, angle=0]{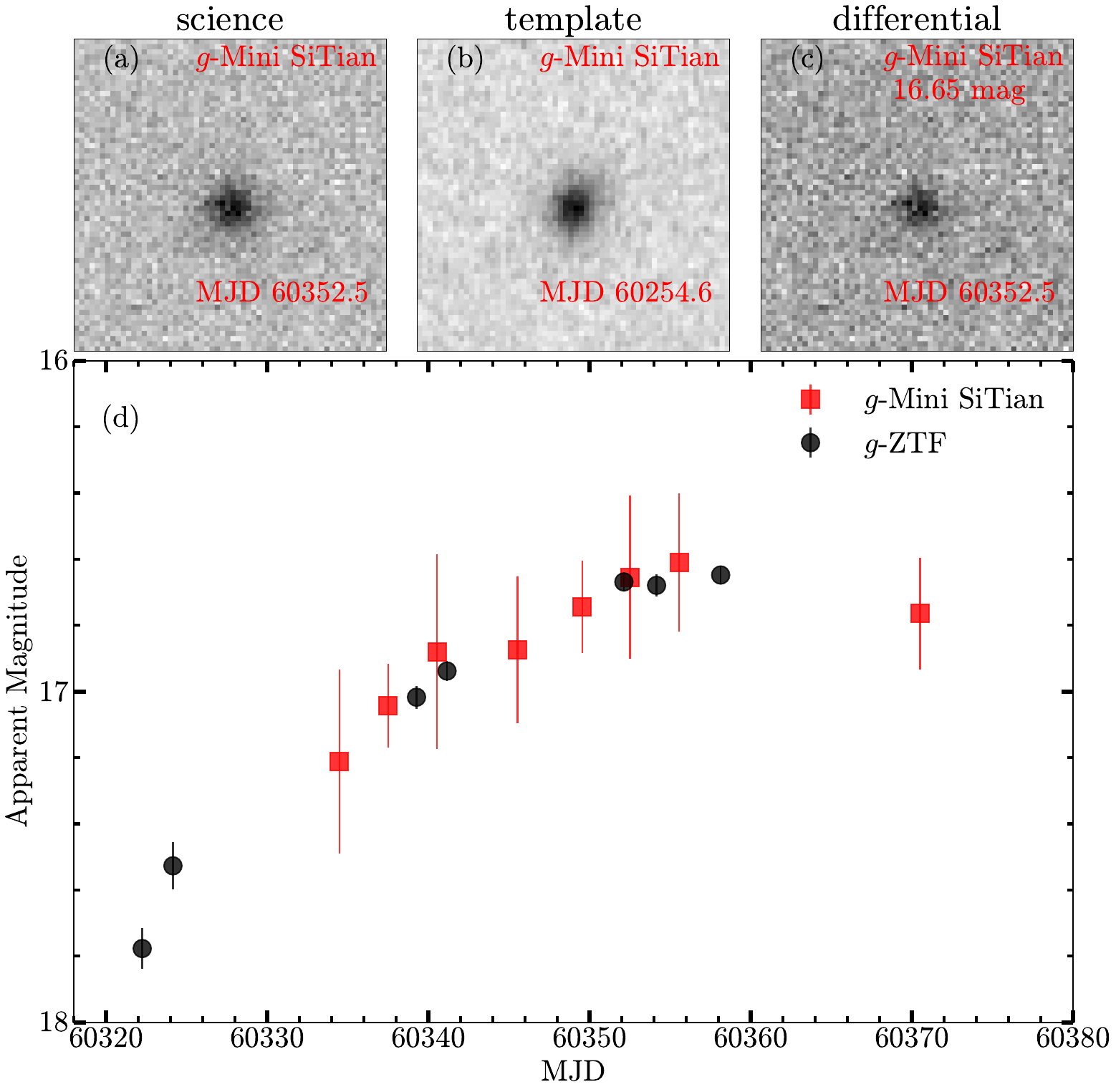}
   \caption{An example of a TDE detected by {Mini-SiTian} in the $g$ band is shown. Panel (a) shows the observation taken on MJD 60352, taken close to the peak of the source. Panel (b) shows the observation on MJD 60254.6, taken before the TDE occurred. Panel (c) shows the differential image. Panel (d) shows the host-galaxy-subtracted light curves of AT2020afhd, directly derived from the differential images obtained by {Mini-SiTian} and ZTF, without correction for Galactic extinction.}
   \label{fig5}
   \end{figure}
   
To verify our results, we search for TDEs in the data collected by {Mini-SiTian} during its first survey run. The first survey commenced in November 2023 and lasted for eight months, concluding in June 2024. Our method is as follows.
Firstly, we divide the celestial sky, with declination ranging from $-20^\circ$ to 90$^\circ$, into equal-area survey regions based on the field of view of the {Mini-SiTian} array. The regions within 10$^\circ$ of the Galactic plane are excluded, due to the high stellar density and significant Galactic dust extinction. We then select the GLADE+ catalog (\citealt{Dalya+etal+2022}), which provides the most extensive galaxy sample necessary for our research, and assign each galaxy to the corresponding celestial region based on its coordinate to determine the galaxy density in each region. Ultimately, we include 87 sky regions with the highest galaxy density. 

Using the method above, we discover one nuclear transient, AT2020afhd, which was classified as a TDE candidate by \citet{Hammerstein+etal+2024}. We show the science, template and differential images of AT2020afhd in Figure \ref{fig5}. To confirm that the transient occurs in the nuclear region, we measure the position of AT2020afhd in the differential image and compare it with the centroid of its host galaxy in the template image. These positions are measured using the barycenter provided by the SEP package (\citealt{Bertin+1996,Bar+2016}). The calculated offset is $0.23''$, smaller than the $0.6''$ threshold used in ZTF (\citealt{Yao+etal+2023}), confirming AT2020afhd as a nuclear transient. Photometric data for {Mini-SiTian} and ZTF are obtained by \textsc{AutoPhOT} (\citealt{Brennan+Fraser+2022}) and the Lasair alert broker\footnote{https://lasair-ztf.lsst.ac.uk/} (\citealt{Smith+etal+2019}), respectively. As shown in Figure \ref{fig5}d, the photometries from {Mini-SiTian} are consistent with those from ZTF. Further analysis of the target will be presented in subsequent works (H.R. Gu et al. 2025). Upon checking the Transient Name Server\footnote{https://www.wis-tns.org/}, AT2020afhd has been currently confirmed as the only TDE that occurred within our survey run and reached the limiting magnitude of {Mini-SiTian}. This demonstrates that {Mini-SiTian} is effective in searching for TDEs and that the prediction of TDE detection in this paper is reliable.
\begin{figure}[tbp]
   \centering
  \includegraphics[width=\textwidth, angle=0]{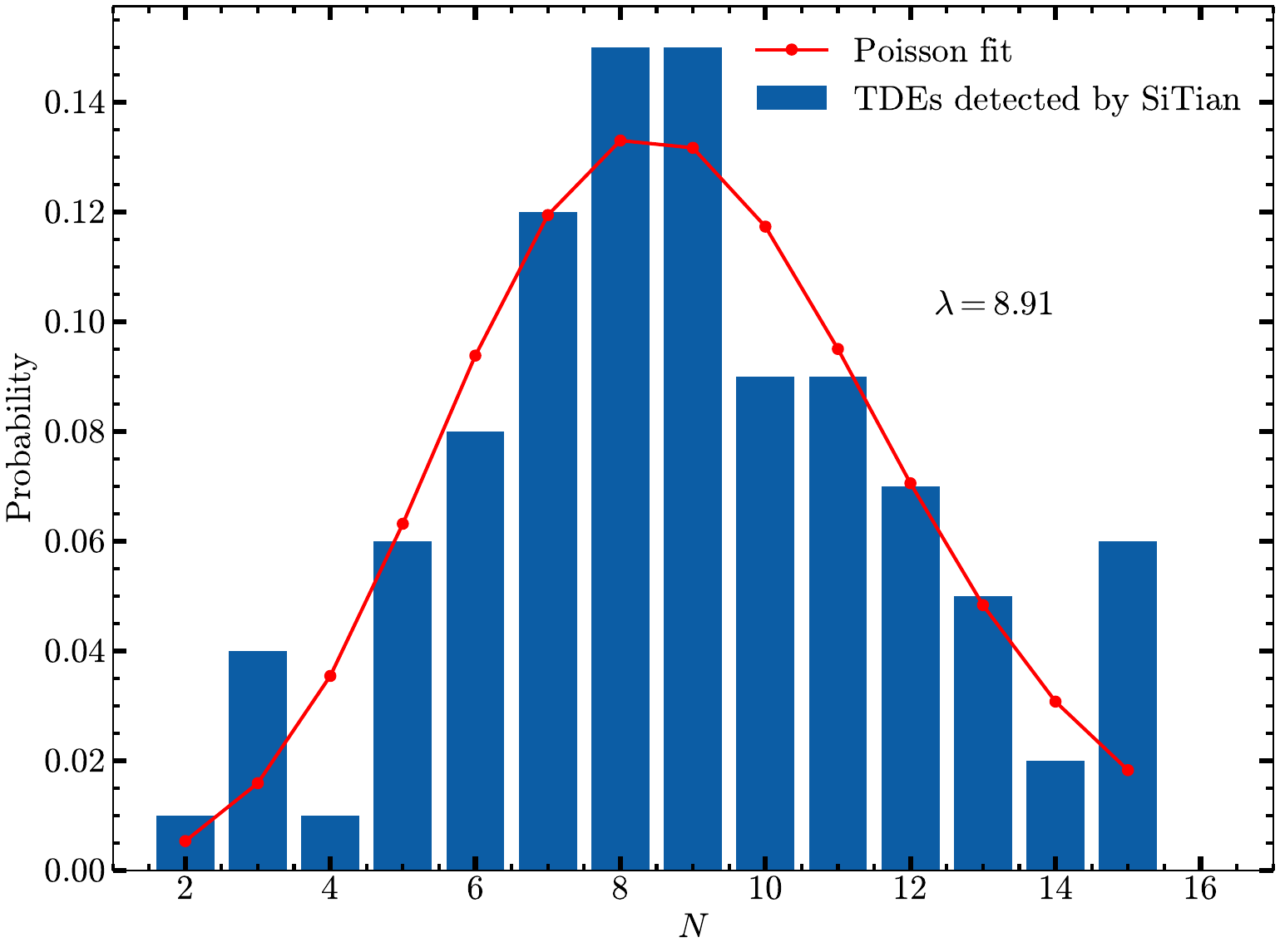}
   \caption{The number distribution of TDEs detected by SiTian for 100 group samples.} 
   \label{fig6}
   \end{figure}
Additionally, we further apply the previous modeling of the {Mini-SiTian} array to the SiTian project, with the limiting magnitude down to 21 mag (\citealt{Liu2021}). For each simulated observation of SiTian, a random number from 0.0 to 1.0 is added to the $g$ and $r$ band limiting magnitudes. This distribution accounts for various observation factors that can reduce the limiting magnitude, such as moonlight pollution, atmospheric extinction, and background noise from host galaxies. Furthermore, for simulations involving the larger FOV of SiTian, described in Section \ref{Mockobs}, $n$ is sampled from a Gaussian distribution with a minimum value of 1 and a maximum value of 6. The distribution has a median and standard deviation both set to 1.

Figure \ref{fig6} shows the distribution of the number of TDEs detected by SiTian. Eventually, the number of TDEs detected by SiTian is approximately {{$9 \pm 3$}} per year over 440 deg$^2$. Extrapolating this to a sky coverage of 10,000 deg$^2$, the estimated TDE rate is $204 \pm 68$ per year, which will significantly increase the sample size of TDEs. Furthermore, SiTian is planned to be installed at the Lenghu Observatory, which has 70 per cent observable nights and median seeing of 0.75 arcsec (\citealt{Deng+etal+2021}). Due to better observation conditions, such as more observable nights and better seeing, the real detection ability of SiTian could be even more powerful.
\section{Discussion}
\label{discussion}
\subsection{Probing the SMBH Mass Distribution}
\label{tdemass}
\begin{figure}[htbp]
   \centering
  \includegraphics[width=\textwidth, angle=0]{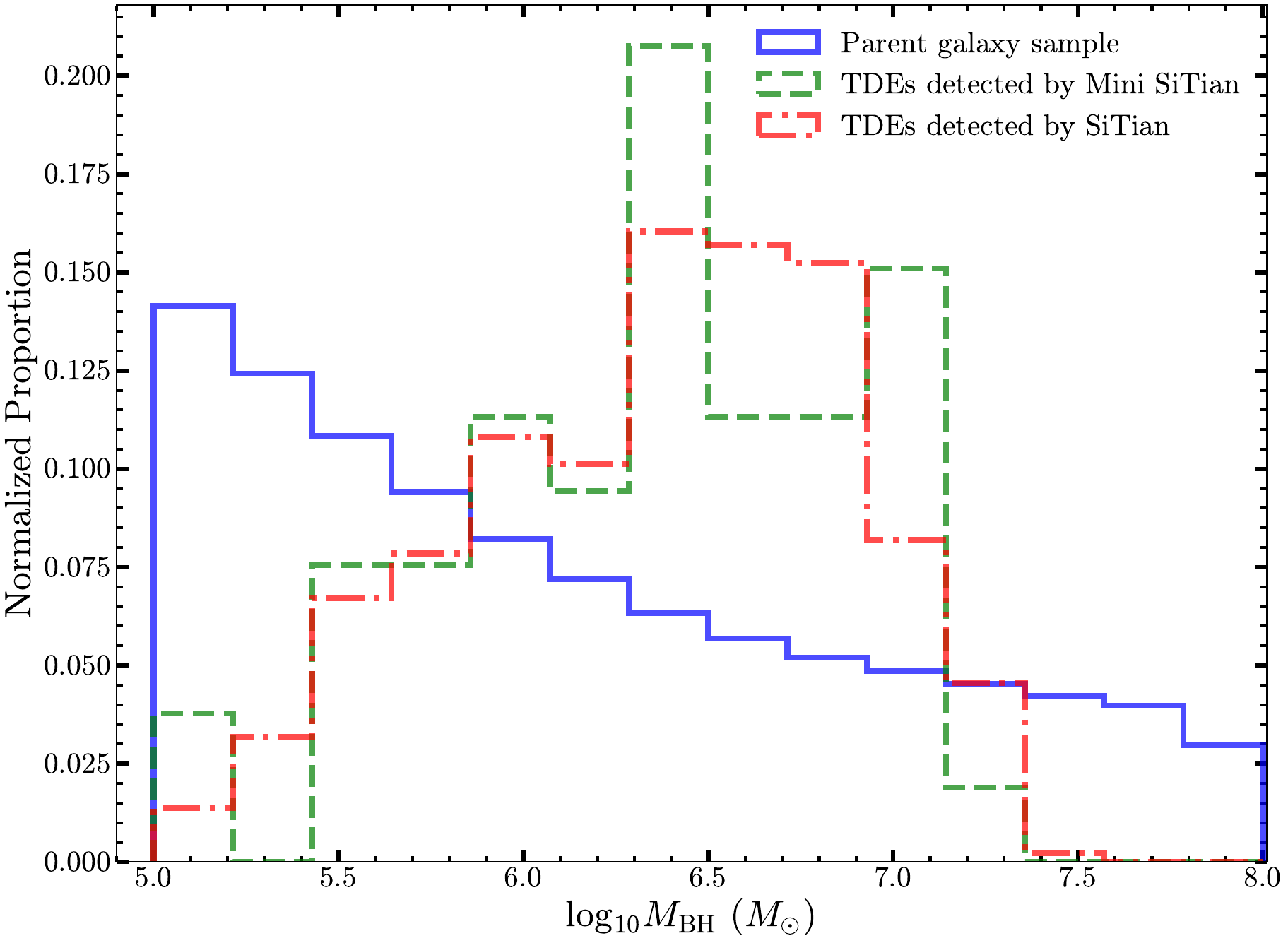}
   \caption{Comparisons of the $M_{\rm BH}$ distribution among three groups: the parent galaxy sample (blue solid line), the TDEs detected by {Mini-SiTian} (green dashed line) and SiTian (red dotted-dashed line). The sample sizes for the parent galaxy sample, the TDEs detected by {Mini-SiTian} and SiTian are 22,605,284, 53 and 891, respectively. For display purposes, the distribution of each sample has been normalized to the same area. } 
   \label{fig7}
   \end{figure}
Figure~\ref{fig7} presents the $M_{\rm BH}$ distribution for the TDEs detected by {Mini-SiTian}, SiTian, and those of our parent galaxy sample. The parent-galaxy group shows a high proportion at the low-mass end ($\leq 10^6 M_{\odot}$), which decreases as the $M_{\rm BH}$ increases. While for the detected-TDEs groups, they display an approximate Gaussian distribution centered around $M_{\rm BH} = 10^{6.3-6.5} M_{\odot}$.
This discrepancy suggests an observational bias: TDEs involving smaller-mass black holes produce dimmer flares that decay more rapidly, making them more challenging to detect.
At the high-mass end ($>10^{7.5} M_{\odot}$), the TDEs rate rapidly decreases, likely due to the fact that there are relatively few galaxies within this mass range, and the tendency of more massive non-spinning black holes to swallow solar-type stars whole without disrupting them.

In fact, as TDEs could reveal the presence of quiescent SMBHs, it provides a unique method for measuring the mass of black holes, independent of the classical $M_{\rm BH}-\sigma$ relation. As mentioned in Section~\ref{sample}, using the {\tt MOSFiT} model along with observational data and theoretical frameworks, we can estimate the mass of SMBHs that produce TDEs. This method is adaptable to various observational scenarios, including cases with robust data sampling near the peak of the event, good data sampling across different wavelengths, or even situations with excellent sampling during the decaying phase but lacking peak observation (\citealt{Mockler2019}).
However, despite the relatively low statistical error—typically within 0.1 dex and only rarely reaching 0.3 dex in determining the mass of SMBHs using this method (\citealt{Mockler2019})—the black hole masses derived from the {\tt MOSFiT} modeling are systematically larger than those determined through the $M_{\rm BH}-\sigma$ relation. This discrepancy suggests that further improvements in the modeling are necessary.
With the advent of large-FOV and high-cadence survey telescopes, we have the opportunity to collect a wide variety of TDE light curves, thereby providing a new and more comprehensive mass distribution of quiescent SMBHs. This expanded dataset will help enhance our understandings of these enigmatic events, the massive objects that drive them and place constraints on theoretical models.
\subsection{Comparison with the ZTF sample}
\label{real samples}
In this section, we present a comparison between the TDEs detected by {Mini-SiTian} (hereafter MST) and those detected by ZTF \citep{Yao+etal+2023} (hereafter Yao23). The authors applied a set of well-defined criteria to select TDE candidates, including specific cuts to isolate nuclear transients, constraints on peak magnitude ($m_{\rm peak}$), the number of detections, and $g-r$ color criteria. This resulted in a flux-limited, spectroscopically complete sample of 33 TDEs over a three-year period (from October 2018 to September 2021). Among these, 16 objects were selected from the ZTF-I phase (from October 1, 2018, to September 30, 2020) with $m_{\rm peak} < 18.75$, and 17 objects were selected from the first year of the ZTF-II phase (from October 1, 2020, to September 30, 2021) in the $g$ band with $m_{g,\rm peak} < 19.1$. The redshift of these TDEs ranges from 0.0151 to 0.519.
\begin{figure}[tbp]
  \centering
  \begin{subfigure}{0.49\textwidth}
    \centering
    \includegraphics[width=\linewidth]{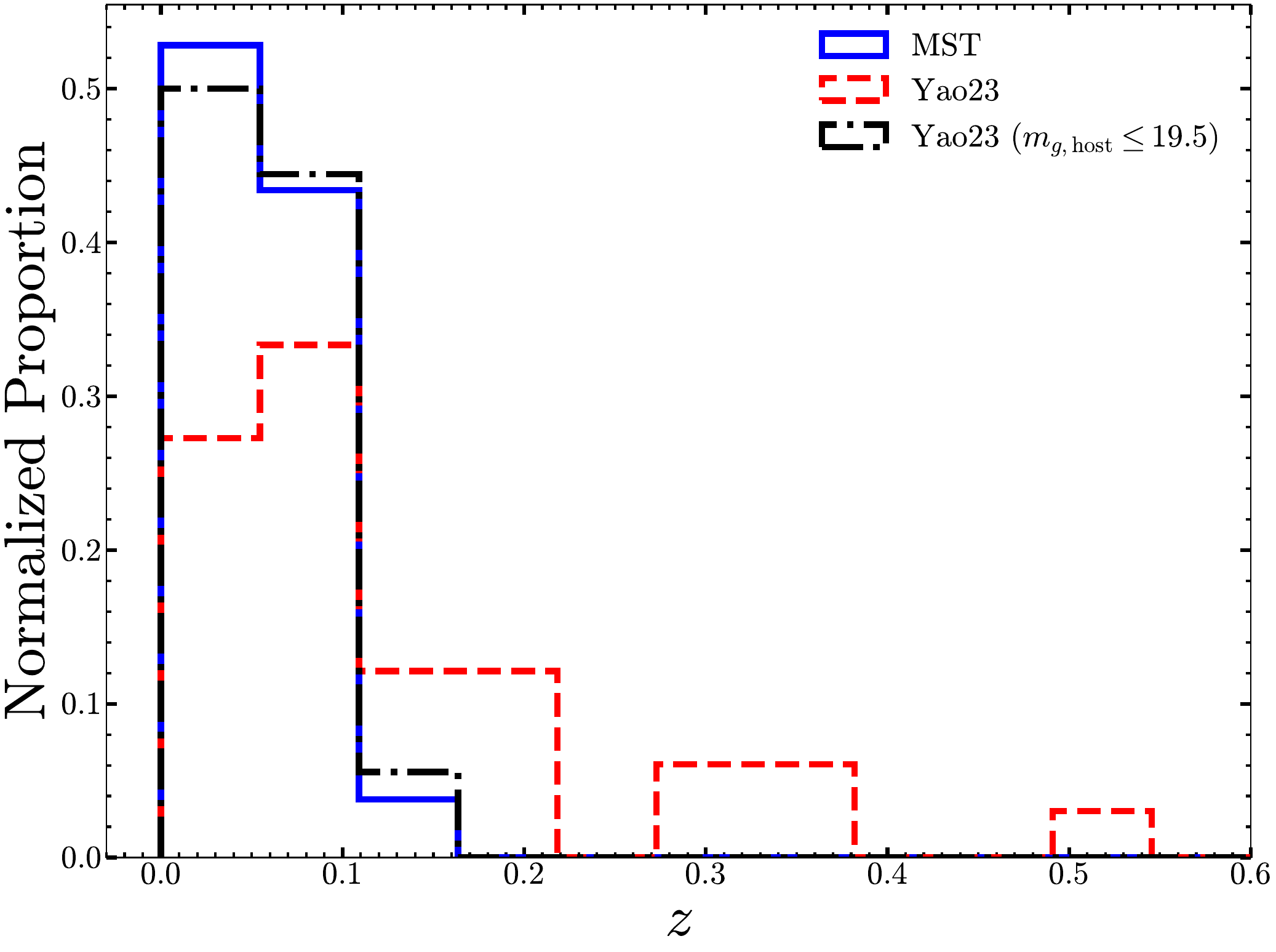}
    \label{fig:z}
  \end{subfigure}
  \hfill
  \begin{subfigure}{0.49\textwidth}
    \centering
    \includegraphics[width=\linewidth]{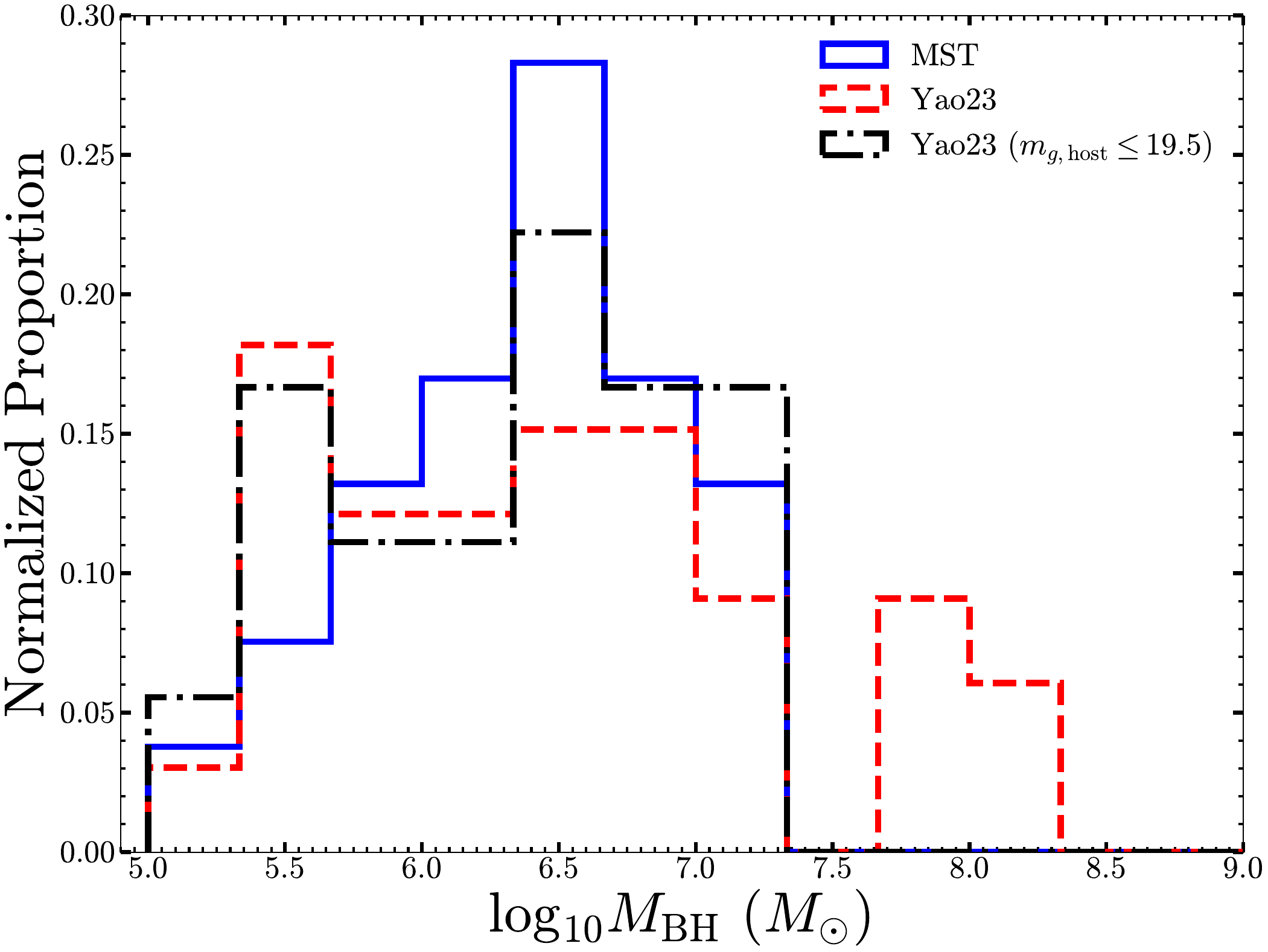}
    \label{fig:mbh}
  \end{subfigure}
  \caption{Distributions of properties of TDEs expected from MST, compared to those discovered by Yao23. Both distributions have been normalized for visualization purposes. The left panel shows the distribution for $z$, while the right panel shows the distribution for $M_{\rm BH}$. Comparisons of the distribution among three samples: MST (blue solid line), Yao23 (red dashed line) and Yao23 $(m_{g,\rm host}\leq 19.5)$ (black dotted-dashed line). The sample sizes for MST, Yao23, Yao23 $(m_{g,\rm host}\leq 19.5)$ are 53, 33 and 18, respectively. K-S tests suggest that MST and Yao23 $(m_{g,\rm host}\leq 19.5)$ distributions are consistent, with the $p$-values of 0.79 for $M_{\rm BH}$ and 0.88 for $z$, respectively.}
  \label{fig8}
\end{figure}
\begin{figure}[htbp]
   \centering
  \includegraphics[width=\textwidth, angle=0]{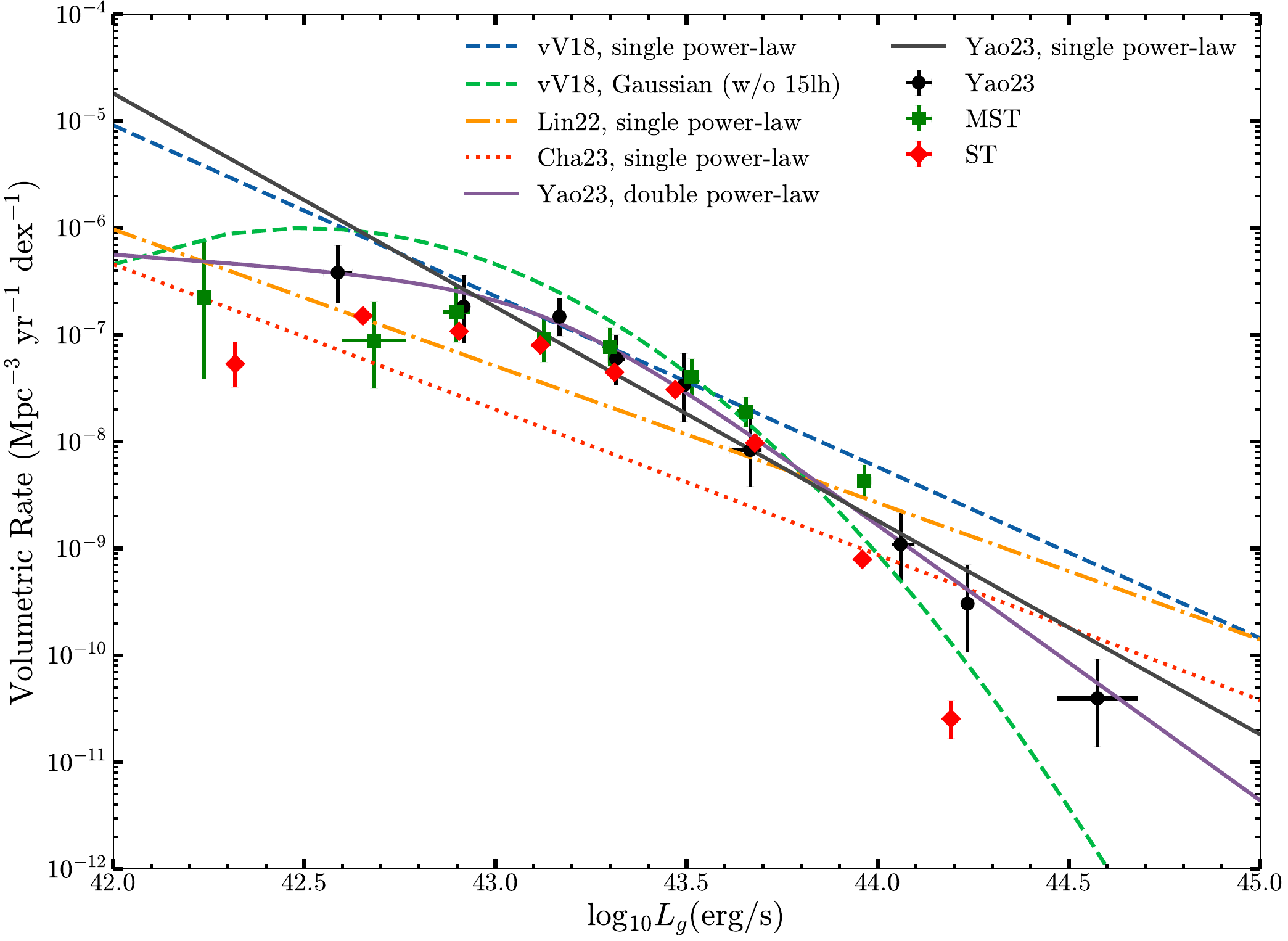}
   \caption{TDE LF in rest-frame $g$-band. The results of Yao23 (black dots), MST (green squares), ST (red diamonds) are plotted, and compared with the TDE LFs from \citet{Velzen2018}, \citet{Lin2022b}, \citet{Charalampopoulos+etal+2023}, and \citet{Yao+etal+2023}. For detected mock TDEs, we perform the volumetric correction, i.e. `1/\textit{V}max' method following \citet{Lin2022b}.} 
   \label{fig9}
   \end{figure}
The distribution of $M_{\rm BH}$ and $z$ of MST and Yao23 is shown in Figure \ref{fig8}. As shown in the figure, the detected TDEs in MST are distributed in the region of $z \sim 0-0.16$ and $M_{\rm BH} \sim 10^{5}-10^{7.3}$ $M_{\odot}$. The median value of black hole mass $M_{\rm BH}=10^{6.40}$ $M_{\odot}$ in MST is close to $M_{\rm BH}=10^{6.44}$ $M_{\odot}$ in Yao23.

To better quantify how well the model distributions align with the observed distributions, we perform Kolmogorov–Smirnov (K-S) tests. For the $M_{\rm BH}$, the $p$-value of 0.31 suggests that the $M_{\rm BH}$ distributions of MST and Yao23 are consistent. However, for the $z$, the $p$-value of 0.002 indicates a significant difference between the two distributions, primarily due to discrepancies above \( z = 0.16 \). The differences in the distributions may be caused by selection bias in the sample.

The sample selection process requires the host galaxy of each TDE to be detected in the template images. However, galaxies at higher $z$ become fainter, making them more difficult to detect in flux-limited surveys. In {Mini-SiTian}, the limiting magnitudes for template images are $m_{g,\rm lim}<19.5$ or $m_{r,\rm lim}<19$, which are significantly shallower than the $m<23$ limit used in Yao23. Consequently, this limits the $z$ range over which we can detect TDEs. To evaluate the impact of this limitation, we exclude the galaxies with $m_{g}>19.5$ in Yao23 (hereafter Yao23 $(m_{g,\rm host}\leq 19.5)$). After re-performing the K-S tests, the $p$ values for the distributions of $M_{\rm BH}$ and $z$ in MST and Yao23 are 0.79 and 0.88, respectively. These results indicate that when the depth of the template images is matched, the distributions between MST and Yao23 become consistent.

The differences in the distributions of MST and Yao23 at larger $M_{\rm BH}$ end ($>10^{7.5} M_{\odot}$) is also attributed to the selection bias in the samples. As shown in Figure \ref{fig8}, five high redshift ($z\geq0.2$) TDEs in Yao23 have $M_{\rm BH}\geq 10^{7.9} M_{\odot}$, which are close to or exceed our Hills mass\footnote{In Section \ref{sample}, using Equation~(\ref{eq1}), we calculate $M_{\rm Hills}=9\times 10^7 M_{\odot}$ for a solar-mass star, which is slightly lower than $M_{\rm Hills}=1.1\times 10^8 M_{\odot}$ obtained by \citet{Yao+etal+2023} using a different equation: $M_{\rm Hills}=1.1\times10^8 M_{\odot}(\frac{M_{\ast}}{M_{\odot}})^{-1/2}(\frac{R_\ast}{R_{\odot}})^{3/2}$.}, making them undetectable in our model. \citet{Yao+etal+2023} attributes these large $M_{\rm BH}$ ($\sim 10^{8.23} M_{\odot}$) to the disruption of evolved stars by relatively large black holes. Modeling such type of disruption is beyond our current work.

To further examine the reliability of our results, we calculate the luminosity function (LF) of detected TDEs using the `1/\textit{V}max' method described in \citet{Lin2022b} and compare it with known optical TDE LFs. As shown in Figure \ref{fig9}, it can be seen that MST, SiTian (represented by ST) and Yao23 exhibit a similar trend:

At the low-luminosity end ($L_{g}\leq10^{43}$ $\rm erg/s$), the LFs for $L_{g}$ flatten. Given a bolometric correction of $\sim$ 10, it corresponds to the Eddington luminosity for BHs with $M_{\rm BH}\leq10^6$ $M_{\odot}$. This behavior may be consistent with an Eddington limited emission scenario.

While at the high-luminosity end ($L_{g}\geq10^{44}$ $\rm erg/s$), there is a noticeable suppression of volumetric rate. This corresponds to the Eddington limit for BHs with $M_{\rm BH}\geq10^7$ $M_{\odot}$. This is due to the suppression of TDE rate when $M_{\rm BH}$ is close to the Hills mass, as shown in Figure \ref{fig7}.
\subsection{Notes on the predicted detection rate}
\label{notes}
Accurately estimating the number of TDEs detected by a survey is a complex task, with several factors that introduce uncertainties are not fully accounted for in our predictions. For example, our simulations do not account for TDE light curves that deviate from a simple power-law decay, such as UV/optical re-brightening or flattening, as observed in cases like AT2022dbl (\citealt{Lin2024}), AT 2018fyk (\citealt{Wevers+etal+2019}), or iPTF15af (\citealt{Blagorodnova+etal+2019}). Such complexities in light curve have not yet been included in {\tt MOSFiT}.

TDEs normally occur at the centers of their host galaxies. Detecting such transients involves additional data processing challenges. The Poisson noise from the host galaxy, along with artifacts from inaccurate image subtraction (\citealt{Hu+etal+2022}) or astrometric alignment issues, can significantly reduce the actual detection efficiency in surveys.

Weather conditions are complex and can affect both the limiting magnitude and the sampling of light curves, thereby reducing the actual detection rate. Additionally, moonlight and airmass can also affect detection limits. To minimize these effects, it is crucial to design the survey strategy to keep the moon at a large angular distance. These additional factors should be considered in future work.

Therefore, a comprehensive survey strategy for {Mini-SiTian} needs to be developed by considering various factors, including the scientific objectives, observing conditions at the Xinglong Observatory, and the operational status of {Mini-SiTian}.

\section{Summary}
\label{summary}
The {Mini-SiTian} array serves as the pathfinder for the wide-field, fast, and deep SiTian project. To evaluate its survey capability, we conduct mock observations to search for TDEs using the {Mini-SiTian} array. Our main results are summarized as follows:

(i) We estimate that {Mini-SiTian} will discover $0.53\pm0.73$ TDEs per year, depending on the limiting magnitude. This estimation requires that at least 10 data points at $g$ and $r$ bands are collected with a cadence of every 3 days over an area of 440 deg$^2$. 
We have discovered one nuclear transient, AT2020afhd, during the first test run of {Mini-SiTian}.

(ii) The SiTian project is predicted to discover up to $204 \pm 68$ TDEs per year, at least ten times the number detected above the limiting magnitude. This will significantly enlarge the TDE sample, improve the statistical understanding of their properties, and enhance our overall knowledge of various transients.

(iii) Our mock observation yields TDEs up to $z \sim 0.16$ for {Mini-SiTian}. The $M_{\rm BH}$ distributions of mock observations and the LF roughly agree with the real optical TDE sample.

\begin{acknowledgements}
We thank the anonymous referee for an extremely quick response and for providing valuable comments, which helped to improve the manuscript. We also thank Zheyu Lin for very useful help and suggestions. We also acknowledge the support of the staff at the Xinglong Observatory. This work is supported by National Key R\&D Program of China (grant No. 2023YFA1609700). This work is supported by the National Natural Science Foundation of China (NSFC; grant Nos. 12090040, 12090041, 12403022 and 12273057). This work is supported by the Strategic Priority Research Program of Chinese Academy of Sciences (grant Nos. XDB0550000, XDB0550100 and XDB0550102). This research is supported by National Key R\&D Program of China (grant No. 2023YFA1608304). Y.H acknowledges the supports of NSFC Nos. of 12422303 and 12261141690, the National Key Basic R\&D Program of China via 2023YFA1608303 and the Strategic Priority Research
Program of the Chinese Academy of Sciences (XDB0550103).
K.X. acknowledges the supports of the NSFC grant No. 12403024, the Postdoctoral Fellowship Program of CPSF under Grant Number GZB20240731, the Young Data Scientist Project of the National Astronomical Data Center, and the China Post-doctoral Science Foundation No. 2023M743447.

The SiTian project is a next-generation, large-scale time-domain survey designed to build an array of {over} 60 optical telescopes, primarily located at observatory sites in China. This array will enable single-exposure observations of the entire {northern hemisphere night sky} with a cadence of only 30-minute, capturing true color ($gri$) time-series data down to about 21 mag. This project is proposed and led by the National Astronomical Observatories, Chinese Academy of Sciences (NAOC). As the pathfinder for the SiTian project, the Mini-SiTian project utilizes an array of three 30 cm telescopes to simulate a single node of the full SiTian array. The Mini-SiTian has begun its survey since November 2022. The SiTian and Mini-SiTian have been supported from the Strategic Pioneer Program of the Astronomy Large-Scale Scientific Facility, Chinese Academy of Sciences and the Science and Education Integration Funding of University of Chinese Academy of Sciences.
\end{acknowledgements}
%% you can type \apj for ApJ, \aap for A&A, \apss for Ap&SS, etc. Please consult
%% the macro chjaa.cls. You can also find them in aasguide.tex (AASTeX for ApJ, AJ, PASP)
%% Please follow the format of ChJAA's reference list
\bibliographystyle{raa}
\bibliography{ms2024-0357.bib}

\label{lastpage}

\end{document}